\documentclass[reprint,amsmath,amssymb,prl,aps,showkeys,superscriptaddress]{revtex4-2}

\usepackage{graphicx}
\usepackage{subfigure}
\usepackage{dcolumn}
\usepackage{bm}
\usepackage{color}
\usepackage{array,multirow}
\usepackage[colorlinks=true,citecolor=blue,linkcolor=blue,urlcolor=blue,bookmarks=true]{hyperref}
\usepackage[sort&compress]{natbib}
\usepackage{ulem}
\renewcommand{\emph}[1]{\textit{#1}}

\newcommand{\sref}[2]{\ref{#1}\hyperref[#1]{#2}}
\usepackage{soul}
\usepackage[table]{xcolor}
\begin{document}
\preprint{APS/123-QED}
\title{\texorpdfstring{Spin-Reorientation-Driven Linear Magnetoelectric Effect in Topological Antiferromagnet Cu$_3$TeO$_6$}{Spin-Reorientation-Driven Linear Magnetoelectric Effect in Topological Antiferromagnet Cu3TeO6}}
\author{Virna Kisi\v{c}ek}
\email{vkisicek@ifs.hr}
\affiliation{Institute of Physics, Bijeni\v{c}ka cesta 46, 10 000 Zagreb, Croatia}
\affiliation{Faculty of Physics, University of Rijeka, Radmile Matej\v{c}i\'{c} 2, 51 000 Rijeka, Croatia}
\author{Damir Dominko}
\affiliation{Institute of Physics, Bijeni\v{c}ka cesta 46, 10 000 Zagreb, Croatia}
\author{Matija \v{C}ulo}
\affiliation{Institute of Physics, Bijeni\v{c}ka cesta 46, 10 000 Zagreb, Croatia}
\author{\v{Z}eljko Rapljenovi\'{c}}
\affiliation{Institute of Physics, Bijeni\v{c}ka cesta 46, 10 000 Zagreb, Croatia}
\author{Marko Kuve\v{z}di\'{c}}
\affiliation{Department of Physics, Faculty of Science, University of Zagreb, Bijeni\v{c}ka cesta 32, 10 000 Zagreb, Croatia}
\author{Martina Dragi\v{c}evi\'{c}}
\affiliation{Institute of Physics, Bijeni\v{c}ka cesta 46, 10 000 Zagreb, Croatia}
\author{Helmuth Berger}
\affiliation{Institute of Physics, Ecole Polytechnique F\'{e}d\'{e}rale de Lausanne (EPFL), CH-1015 Lausanne, Switzerland}
\author{Xavier Rocquefelte}
\affiliation{Univ Rennes, CNRS, ISCR (Institut des Sciences Chimiques de Rennes) UMR 6226, F-35000 Rennes, France}
\author{Mirta Herak}
\email{mirta@ifs.hr}
\affiliation{Institute of Physics, Bijeni\v{c}ka cesta 46, 10 000 Zagreb, Croatia}
\author{Tomislav Ivek}
\affiliation{Institute of Physics, Bijeni\v{c}ka cesta 46, 10 000 Zagreb, Croatia}

\date{\today}
%
\begin{abstract}
The search for new materials for energy-efficient electronic devices has gained unprecedented importance. Among the various classes of magnetic materials driving this search are antiferromagnets, magnetoelectrics, and systems with topological spin excitations. Cu$_3$TeO$_6$ is a material that belongs to all three of these classes. Combining static electric polarization and magnetic torque measurements with phenomenological simulations we demonstrate that magnetic-field-induced spin reorientation needs to be taken into account to understand the linear magnetoelectric (ME) effect in Cu$_3$TeO$_6$. Our calculations reveal that the magnetic field pushes the system from the nonpolar ground state to the polar magnetic structures. However, nonpolar structures only weakly differing from the obtained polar ones exist due to the weak effect that the field-induced breaking of some symmetries has on the calculated structures. Among those symmetries is the $PT$ ($\overline{1}'$) symmetry, preserved for Dirac points found in Cu$_3$TeO$_6$. Our findings establish Cu$_3$TeO$_6$ as a promising playground to study the interplay of spintronics-related phenomena.
\end{abstract}
%
\maketitle
\begin{figure}[tb]
\centering
\begin{centering}
    \setlength{\belowcaptionskip}{-5pt}
    \includegraphics[width=0.81\columnwidth]{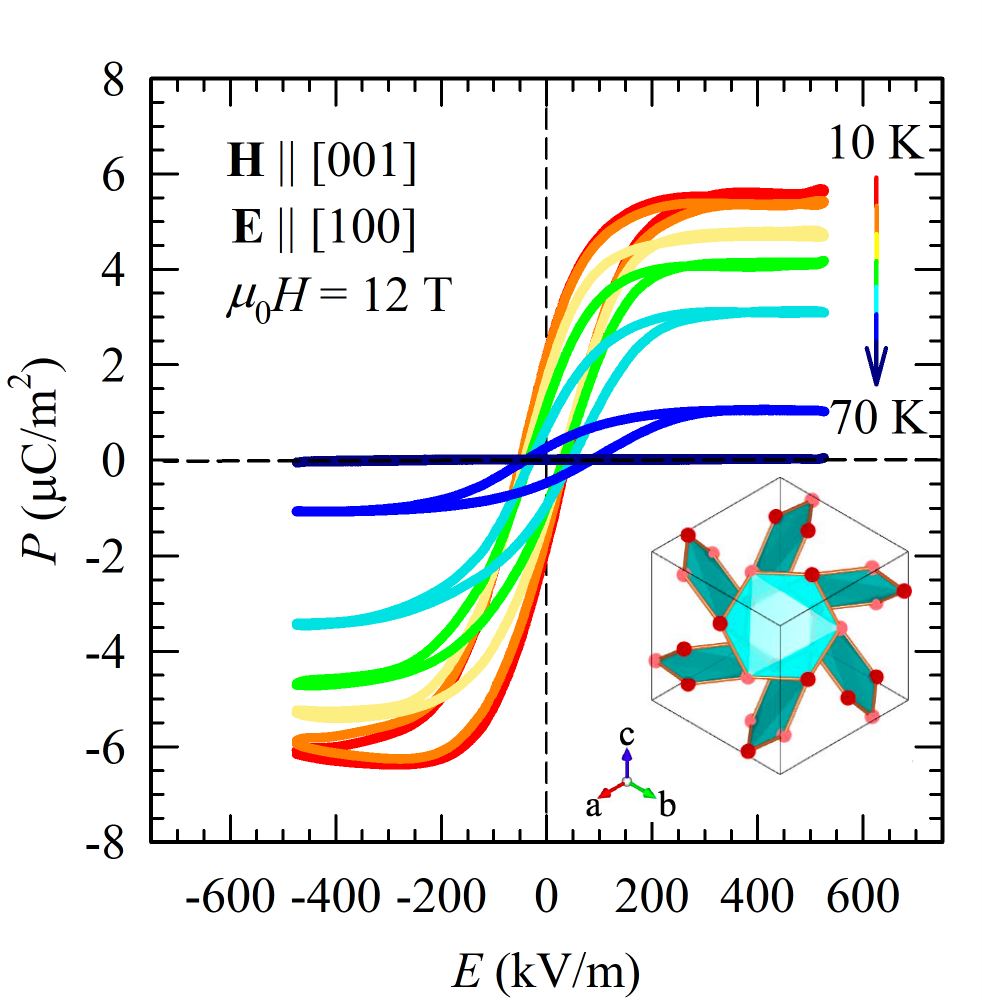}
    \centering
    \caption{Ferroelectric contribution to polarization as a function of applied electric field and temperature measured for $\textbf{H} \| [001]$ and  $\textbf{E} \| [100]$ in applied external magnetic field of 12\,T and temperature range from 10 to 70\,K in steps of 10\,K. Inset: GS AFM structure plotted in unit cell for one of the domains. See supplemental material (SM) \cite{suppl-mater} for details.}
	\label{fig1}
\end{centering}
\end{figure}
\begin{figure*}[t]
\setlength{\belowcaptionskip}{-5pt}
\includegraphics[width=0.98\textwidth]{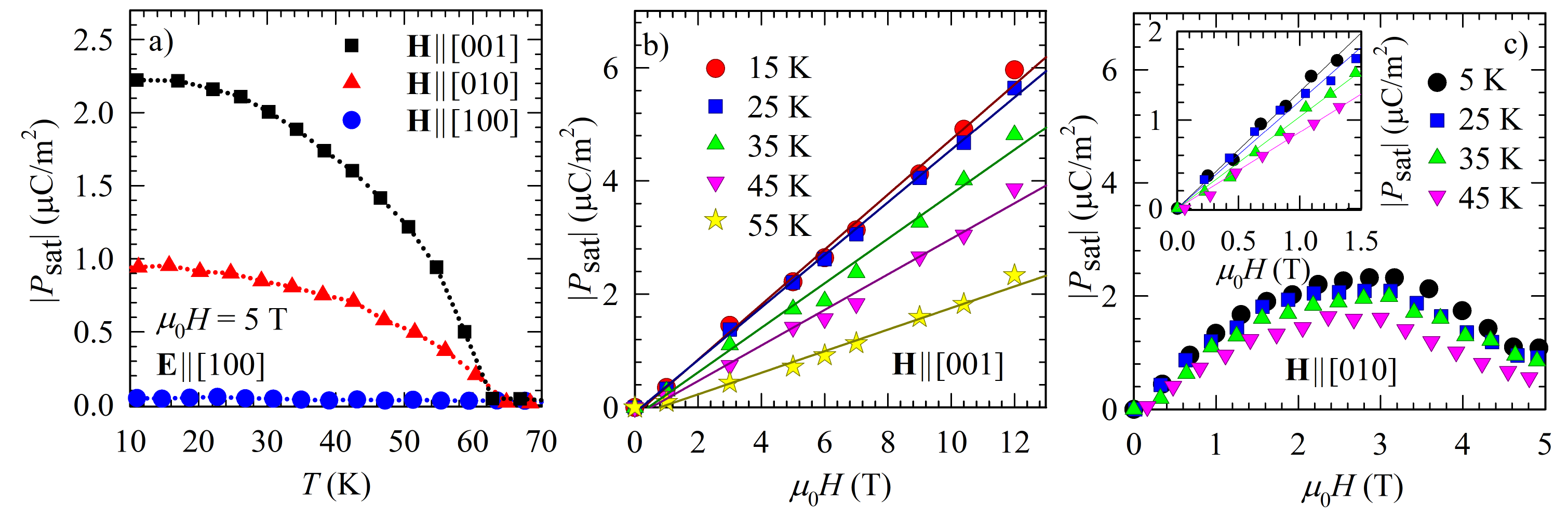}
\vspace{-10pt}
\caption{a) Saturation polarization $P_{\textup{sat}}$ dependence on temperature for three different orientations of the applied external magnetic field $\mu_0 H$\,=\,$5$\,T with respect to the electric field $\textbf{E} \| [100]$. b) $P_{\textup{sat}}$ as a function of magnetic field (0\,--\,12\,T) and temperature (15\,--\,55\,K) measured for $\textbf{H} \| [001]$ (symbols). Linear fit (lines) gives ME coupling coefficients. c) $P_{\textup{sat}}$ as a function of magnetic field (0\,--\,5\,T) measured for $\textbf{H} \| [010]$ (symbols) in the temperature range from 5 to 45\,K. Inset: linear fit (lines) for $\mu_0 H \le 1.5$\,T gives ME coupling coefficients at different temperatures.}
\label{fig2}
\end{figure*}
\begin{figure*}[tb]
\centering
    \setlength{\belowcaptionskip}{-5pt}
    \includegraphics[width=0.98\textwidth]{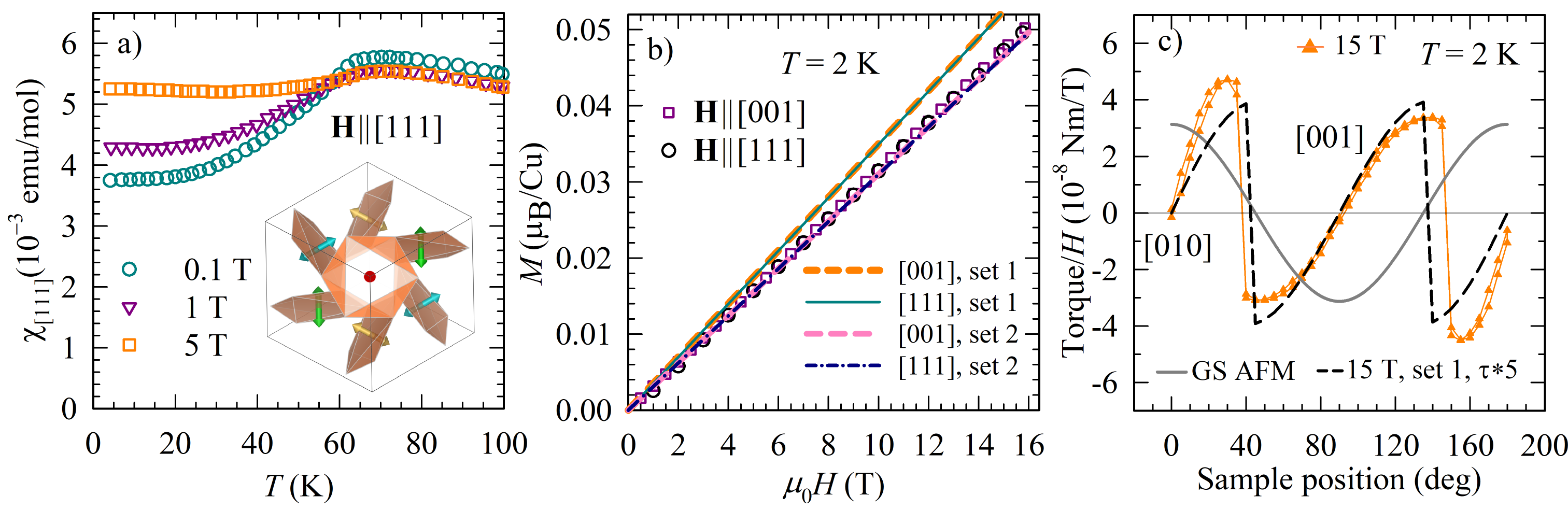}
    \caption{a) Temperature dependence of magnetic susceptibility in different applied fields for $\mathbf{H} \| [111]$. Inset: schematic of AFM domains in the GS \cite{Herak-2005,suppl-mater,Vesta}. b) Field dependence of magnetization at $T = 2$\,K for $\mathbf{H} \| [001]$ and $\mathbf{H} \| [111]$ (symbols) and calculations (lines). Set 1 and 2 represent the choice of superexchange parameters from Ref. \onlinecite{Yao} and \onlinecite{Wang-2019}, respectively. c) The angular dependence of magnetic torque $\boldsymbol{\tau}$ measured at $T=2$~K in the $([010],[001])$ plane (symbols). $\boldsymbol{\tau}$ calculated under the assumption of GS AFM structure (solid line) is compared to $\boldsymbol{\tau}$ calculated from free energy \eqref{eq:freeenergy} (dashed line). The amplitude of the calculated torque is multiplied by 5 for the latter case.}
	\label{fig3}		
\end{figure*}
\begin{figure}[tb]
\centering
   \setlength{\belowcaptionskip}{-5pt}
    \includegraphics[width=\columnwidth]{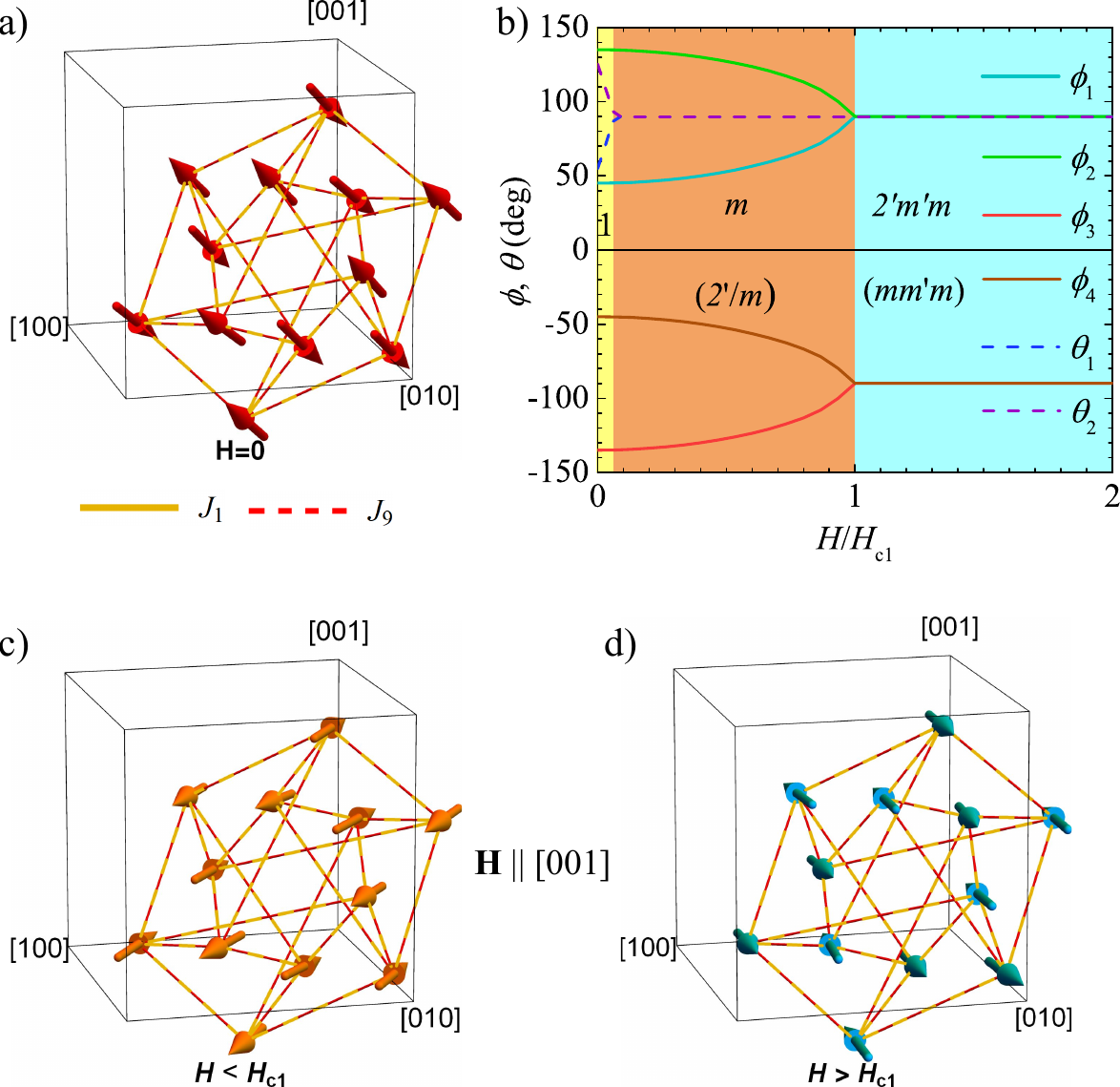}
    \caption{a) Calculated GS magnetic structure for one domain. b) Magnetic phase diagram obtained for $\mathbf{H}\|[001]$ and $D>0$. $\phi$ and $\theta$ are azimuthal and polar angles of the calculated N\'{e}el vectors. The phases correspond to magnetic structures in c) and d) with MPGs $m$ and $2'm'm$, respectively. c) Calculated magnetic structure for $H_{c0}$\,$<$\,$H$\,$<$\,$H_{c1}$ and d) $H \geq H_{c1}$.}
 		\label{fig4}
\end{figure}

%
\indent Antiferromagnetic (AFM) materials are currently a focus of materials research thanks to the fields of spintronics \cite{AFMspintronics-2018} and magnonics \cite{2021_Magnonics_Roadmap}. Large exchange interactions between spins in AFM materials yield spin dynamics at terahertz frequencies and no stray fields make them a natural choice for potential applications in ultrafast spintronic devices \cite{Song-2018,Jungwirth-2016,*Jungwirth-2018}. The emerging field of topological magnets has a promising potential in information technology. Owing to the robustness against many perturbations these materials offer a route to the more energy-efficient memory devices, while magnetic excitations (e.g.\ magnons) could be used to transform and process the information \cite{Malki-2020,*McClarty-2022,*Bernevig-2022}. Topological AFM materials within spintronics, promise new applications in the future technologies \cite{Smejkal-2018,*Bonbien-2022}.\\
\indent Recent interest in cuprates has shown this family offers a vast playground of exotic ground states and phenomena, such as high-temperature superconductivity, magnetic insulating state, layered crystal structure and strong coupling between spins, charge and orbital degrees of freedom \cite{Khomskii-2022,Norman-2018,vuletic2006,Proust,Dong-2019}. Such couplings can lead to ME effect, i.e.\ appearance of polarization $\mathbf{P}_{i}$\,=\,$\alpha_{ij}\mathbf{H}_j$ or magnetization $\mu_0 \mathbf{M}_{j}$\,=\,$\alpha_{ji}\mathbf{E}_i$ in a magnetic or electric field, respectively, as defined by ME tensor $\alpha_{ij}$ \cite{fiebig,Spaldin-2019,Thoele-2020,Rivera,Schmid-2008}. Magnetoelectrics open a way to possible applications in data processing and data storage \cite{Ortega,Spaldin-2020,Spaldin-2019}, but also in the fundamental understanding concerning the opposite requirements for the $d$-orbital occupancy for the cross-coupling to emerge \cite{Hill-2000,Dong-2019,Spaldin-2020}.\\
\indent In terms of symmetry analysis, the ME effect vanishes in systems with one of the space inversion \emph{P}\,($\overline{1}$) or time reversal \emph{T}\,($1$') symmetries, while it is permitted in systems with $PT$ symmetry ($\overline{1}'$). Linear ME coupling in these systems may be generated from the well-known spin-driven ferroelectricity mechanisms (exchange striction mechanism, inverse Dzyaloshinskii-Moriya interaction and spin-dependent \textit{p-d} hybridization) \cite{Tokura-2014,Dong-2015,Fiebig-2016}, unconventional magnetic ordering \cite{Arima} as well as a few symmetrically distinct multipole moments \cite{Schmid-2008,Fiebig-2016}. \\
\indent Cu$_3$TeO$_6$ is a tellurium-based cuprate \cite{Norman-2018} which crystallizes in a cubic $Ia\overline{3}$ space group \cite{Falck}. The Cu$^{2+}$ ions carry spin $S$\,=\,$1/2$ and lead to an AFM ordered ground state (GS) below the N\'{e}el temperature $T_N \approx 62$\,K, described by trigonal magnetic space group $R\overline{3}'$ \cite{Herak-2005}. The first-nearest-neighbor (NN) interaction between the spins defines a 3D network of corner-sharing hexagons (inset of Fig.\ \sref{fig1}{a}). The spins in the AFM state are almost collinear and aligned along one of the $\left< 111\right>$ directions of the cubic unit cell \cite{Herak-2005,Herak-2011,Li} resulting in the presence of multiple AFM domains \cite{suppl-mater}. Optical measurements have revealed the magnetoelastic effect deep in the AFM state induced by the spin-phonon coupling \cite{Caimi,Choi}. Inelastic neutron scattering (INS) \cite{Bao-2019,Yao} confirmed Heisenberg spin model predictions \cite{Li} of topological Dirac and nodal line magnons with $PT$ symmetry preserved. The same technique, in combination with thermodynamic studies, revealed magnon-polaron mode representing the collective excitations resulting from the magnon-phonon coupling \cite{Bao-2020, Chen-2022}. Moreover, a unique magnetic lattice of Cu$_3$TeO$_6$ was proposed to be at the origin of the spin gap observed in the nuclear magnetic resonance measurements \cite{Baek-2021}.\\
\indent In this Letter we report the previously unobserved influence of the magnetic-field-induced spin reorientation and related symmetry on the linear magnetoelectric effect in Cu$_3$TeO$_6$, establishing this material as a playground to study the interplay of spintronics-related phenomena.\\
\indent High-quality single crystals of Cu$_3$TeO$_6$ were grown using HBr chemical transport method in sealed quartz tubes \cite{Herak-2005} and characterized using an X-ray diffractometer at room temperature. Ferroelectric (FE) polarization hysteresis loops were obtained using a homemade Sawyer-Tower-type virtual ground setup \cite{Prume-2004} as described in \cite{Dragicevic-2021} with a frequency set to 77\,Hz in a quasi-static electric field up to 500\,kV/m \cite{suppl-mater}. The magnetic properties of Cu$_3$TeO$_6$ were studied using Quantum Design (QD) Magnetic Properties Measurement System (MPMS3) magnetometer and vibrating sample magnetometer and torque magnetometer on QD Physical Properties Measurement System.\\
%
%
%
\indent Our quasi-static electric polarization measurements show that FE polarization is induced by magnetic field in AFM state. $P(E)$ hysteresis loops (Fig.\ \ref{fig1}) measured in 12\,T for $\textbf{H} \| [001]$ and $\textbf{E} \| [100]$ in the temperature range from 10 to 70\,K below $T_N$ are slightly biased and saturated above 250\,kV/m, while above $T_N$ they vanish. Similar behavior is observed for $\textbf{H} \| [010]$ and $\textbf{E} \| [100]$ \cite{suppl-mater}.\\
\indent Saturation polarization $P_{sat}$ dependence on temperature and magnetic field measured for $\textbf{E} \| [100]$ is shown in Fig.\ \ref{fig2}. It shows non-zero values for $T\leq T_N$ and $\mathbf{E} \perp \mathbf{H}$ (Fig.\ \sref{fig2}{a}). $P_{sat}$ dependence on magnetic field $\mathbf{H}\|[001]$ is linear for temperatures ranging from 15 to 55~K (Fig.\ \sref{fig2}{b}). Interestingly, for $\mathbf{H}\|[010]$ and $\mathbf{E}\| [100]$, $P_{sat}$ increases with the field, reaching a maximum at $\mu_0 H=2.5-3$~T and then decreases (Fig.\ \sref{fig2}{c}). The ME coupling coefficients obtained at 5\,K are $\alpha_{ac}= 0.61$\,ps/m, and $\alpha_{ab}= 1.66$\,ps/m (in the low-field region) for $\mathbf{H}\|[001]$ and $\mathbf{H}\|[010]$, respectively. Here, we use cubic crystal coordinate system $(a,b,c)$ with $a$\,=\,$[100]$, $b$\,=\,$[010]$ and $c$\,=\,$[001]$ for notation.\\
%
%
%
%
\indent The temperature dependence of magnetic susceptibility $\chi$ measured in several different magnetic field values for $\textbf{H}\|[111]$ (Fig.\ \sref{fig3}{a}), with a visible kink at $T_N \approx 62$\,K, is consistent with the previous findings \cite{Herak-2005, He-2014, Choi, Bao-2019}. Below $T_N$ the susceptibility increases as the field increases, which is typical for AFM materials with multiple orientational domains where the spin reorientation is taking place in the applied magnetic fields \cite{Zhao-2017}. No difference was observed between the zero-field-cooled and field-cooled curves. \\
\indent The field dependence of magnetization $M$ for $\mathbf{H}\|[001]$ and $\mathbf{H}\|[111]$ measured at 2\,K, is presented in Fig.\ \sref{fig3}{b}. In the entire range, $M$ seems to be linear in $H$ and isotropic with mild nonlinearity observed for $\mathbf{H}\|[111]$ \cite{suppl-mater,Tang-2023}.\\ 
\indent In Fig.\ \sref{fig3}{c} we plot the angular dependence of magnetic torque $\boldsymbol{\tau}$ in AFM state measured at 2\,K in the $([010],[001])$ plane in $\mu_0H$\,=\,$15$~T. $\boldsymbol{\tau}$ displays a sharp change of sign for field angles in the vicinity of the $[011]$ direction. Such behavior deviates from the $\tau \propto \sin 2\theta$ dependence expected for AFM with no spin reorientation (solid line in Fig.~\sref{fig3}{c}), and is obtained for $\mu_0 H$ ranging from 1 to 15~T \cite{suppl-mater}.  \\
%
\indent To determine the magnetic structure, we start with the Hamiltonian 
\begin{align}\label{eq:Hamiltonian} \nonumber
	\mathcal{H} &= J_1\sum_{\langle i,j \rangle} \mathbf{S}_i \cdot \mathbf{S}_j + J_9\sum_{\langle i,k \rangle} \mathbf{S}_i \cdot \mathbf{S}_k \;+\\ 
		& +d_{DMI} J_1 \sum_{\langle i,j \rangle} \mathbf{d}_{ij} \cdot (\mathbf{S}_i \times \mathbf{S}_j) -\mathbf{H}\cdot \sum_i \mathbf{\hat{g}}_{i}\cdot \mathbf{S}_i,
\end{align}
where $J_1$ and $J_9$ are the two dominant isotropic interactions between the first-NN ($d_\textrm{Cu--Cu}$\,=\,3.18\,\AA) and ninth-NN ($d_\textrm{Cu-Cu}$\,=\,6.21\,\AA), respectively \cite{Yao, Wang-2019}. Dzyaloshinskii-Moriya interaction (DMI) \cite{Dzyaloshinsky-1958,*Moriya-1960} is introduced between the first-NN. The last term is the Zeeman interaction where $\mathbf{\hat{g}}_{i}$ is the electron \textbf{g} tensor of spin $i$. The orientation of the Dzyaloshinskii-Moriya (DM) unit vector $\mathbf{d}_{ij}$ was obtained from Cu-O-Cu bond geometry \cite{suppl-mater}. $D=d_{DMI} J_1$ is the magnitude of DMI. The direction of the DMI vector is defined by setting $D$\,$>$\,$0$ or $D$\,$<$\,$0$, and it has important consequences for the ME effect. The summation for $J_1$ interaction and DMI goes over first-NN spins and for $J_9$ over ninth-NN spins where each spin has 4 of both NN \cite{Herak-2005,Wang-2019,Yao}.\\
\indent The primitive cell consisting of 12 magnetically inequivalent Cu$^{2+}$ ions fully describes the magnetic structure in Cu$_3$TeO$_6$ \cite{Herak-2005}. We use this primitive cell (Fig.\ \sref{fig4}{a}) in our simulations. Next, we map the interactions $J_1$ and $J_9$ onto the minimal cell by considering the boundary conditions (see SM \cite{suppl-mater}). The resulting magnetic lattice with interactions $J_1$ and $J_9$ mapped onto primitive cell is shown in Fig.\ \sref{fig4}{a}. Interestingly, both interactions mapped onto the same lattice result in effective 4 NN which might be a signature of low-dimensionality hidden in this topologically unique 3D magnetic lattice \cite{Herak-2005,Baek-2021}.\\
%
\indent From the Hamiltonian \ref{eq:Hamiltonian} we write the free energy $\mathcal{F}$
\begin{align}\label{eq:freeenergy}\nonumber
	&\mathcal{F} = \dfrac{k_B}{g_e^2 \cdot \mu_B \cdot 10^{4}}\left[ J_1 \sum_{\langle i,j \rangle} \mathbf{M}_i\cdot \mathbf{M}_j + J_9 \sum_{\langle i,j \rangle} \mathbf{M}_i\cdot \mathbf{M}_j \right.\\ 
 & + \left.  d_{DMI}J_1\sum_{\langle i,j \rangle} \mathbf{d}_{i,j} \left(\mathbf{M}_i\times\mathbf{M}_j\right) \right]-\mathbf{H} \sum_i \mathbf{\hat{g}}_{i} / g_e\mathbf{M}_i,
\end{align}
where $k_B$ is Boltzmann constant, $g_e=2.0023$ is the free electron $g$ value and $\mu_B$ is Bohr magneton. In the minimal cell the summation for both $J_1$ and $J_9$ goes over the same pairs $\langle i,j \rangle$ (see Fig.\ \sref{fig4}{a}). With $S_0$\,=\,$1/2$, and sublattice magnetization for spin $i$ $\mathbf{M}_i$\,=\,$S_0(\sin \theta_i \cos \phi_i, \sin \theta_i \sin \phi_i, \cos \theta_i)$. The calculated $\mathbf{\hat{g}}_{i}\,\forall i$ is given in SM \cite{suppl-mater}. $\theta_i$ and $\phi_i$ are polar and azimuthal coordinates with corresponding Cartesian system $([100], [010], [001])$. The magnitude of the applied magnetic field $H$ is expressed in Tesla (T) units. We performed calculations with two sets of parameters: 1) $J_1= J_9 = 4.8$\,meV and $D=0.1 J_1$, proposed from the INS experiment \cite{Yao}, and 2) $J_1=7.05$~meV, $J_9=3.77$~meV and $D=0.06 J_1$, proposed from theory \cite{Wang-2019}. The free energy, Eq. \eqref{eq:freeenergy}, is minimized using the quasi-Newton method. The resulting magnetic structure is used to calculate the total magnetization and torque and to determine the preserved symmetry elements. In this way, the magnetic point group (MPG) was found in zero and finite magnetic fields \cite{Aroyo-2006,*Aroyo-2011,*Litvin}.\\
\indent The two sets of parameters lead to the same calculated GS, which is shown for one of the 8 AFM domains in Fig.\ \sref{fig4}{a}. An excellent agreement is obtained with neutron diffraction experiment \cite{Herak-2005}. This GS is 8-fold degenerate with 8 AFM domains with dominant spin orientation (easy axis) along $\left< 111 \right>$ directions. The weak canting of spins amounts to $\approx 1^{\circ}-2^{\circ}$, depending on the chosen set of parameters, in good agreement with theory \cite{Wang-2019}. The magnetic point group of the calculated structure is $\overline{3}'$. \\ 
\indent The calculated magnetization per Cu for $\mathbf{H}\|[001]$ and $\mathbf{H}\|[111]$ for two sets of parameters is shown in Fig. \sref{fig3}{b}, where slightly better agreements is observed for the second set \cite{Wang-2019}. The calculated magnetic torque $\tau$ with parameters from Ref. \onlinecite{Yao}, (Fig. \sref{fig3}{c}), captures the angular dependence of the measured curves very well. The sharp sign change of $\tau$ in the vicinity of $[011]$ direction is observed as a signature of the spin reorientation, as well as the correct phase. The obtained amplitude of torque is, however, 5 times smaller than the one in the experiment, signifying that magnetic anisotropy is underestimated in our model. Increasing the DMI to $D=0.3 J_1$ almost reproduces the measured amplitude \cite{suppl-mater}. On the other hand, the torque calculated under the assumption of the GS structure and $\approx 10\%$ domain imbalance (solid line in Fig.~\sref{fig3}{c}) is in stark disagreement with the experiment. \\
\indent The main result of our analysis is the magnetic-field-induced spin reorientation which is captured by a rotation of the N\'{e}el vector $\mathbf{l}$ in applied magnetic field, $\mathbf{l}= (\sum_{i=1}^6 \mathbf{M}_{i,\uparrow} - \sum_{j=1}^6 \mathbf{M}_{j,\downarrow})/(12\,M)$, where we distinguish the moments with opposite main components as $\mathbf{M}_{i,\uparrow}$ and $\mathbf{M}_{j,\downarrow}$. For domain $i$, the direction of $\mathbf{l}_i$ is described by $(\theta_i,\phi_i)$. In Fig. \sref{fig4}{b} we plot the magnetic phase diagram calculated for $\mathbf{H}\|[001]$ for all domains. The spin reorientation takes place as soon as the finite $H$ is applied. Three phases are observed. A phase with MPG 1 is found for $\mu_0 H_{c0} \lesssim 0.04$~T, a field too low to induce a measurable ME effect. We focus on the two other phases separated by the critical field $H_{c1}$. The magnetic structures in $\mathbf{H}\|[001]$ for $H_{c0}$\,$<$$H$\,$<$$H_{c1}$ and $H\geq H_{c1}$ are shown in Figs.\ \sref{fig4}{c} and \sref{fig4}{d} respectively. The MPGs of the calculated structures for $H\geq H_{c1}$ depend on the direction of the DM vector, defined by the sign of $D=d_{DMI} J_1$. $D>0$ gives MPGs in agreement with our polarization measurements. MPGs for $D<0$ can be found in SM. For $H_{c0}$\,$<$$H$\,$<$$H_{c1}$ the MPG is $m$ ($m\perp\mathbf{H}$), and in $H\geq H_{c1}$ it is $2'm'm$. The magnitude of the critical field $\mu_0 H_{c1}$ amounts to $0.75$~T and $0.6$~T for the parameters from Refs. \onlinecite{Yao} and \onlinecite{Wang-2019}, respectively. The $\mu_0 H_{c1}\approx 3$~T suggested from Fig.~\sref{fig2}{c} is reproduced in our model for $D \approx 0.25 J_1$ \cite{suppl-mater}. Alternatively, another source of magnetic anisotropy energy might be needed to fully capture the magnetic anisotropy of this system.\\
%
%
%
\indent In table \ref{tab:symmetry} we list the preserved symmetry operations and the MPGs obtained in our calculations for $D>0$ \cite{Aroyo-2006,*Aroyo-2011,*Litvin}. The MPGs obtained for $H>0$ are polar, in contrast to the nonpolar GS. However, the effect of the field-induced breaking of some symmetries (marked by $\pm$) which would lead to nonpolar MPGs might be too weak to be observable in moderate magnetic fields. We add those nonpolar MPGs in parentheses in table \ref{tab:symmetry} and Fig.~\sref{fig4}{b}. Among those symmetries is $\overline{1}'$.  \\
\setlength{\tabcolsep}{0pt}
\newcolumntype{C}[1]{>{\centering\let\newline\\\arraybackslash\hspace{0pt}}m{#1}}
\renewcommand{\arraystretch}{1.2}
 \definecolor{Gray}{gray}{0.9}
\begin{table}[tb]
	\begin{center}
		\begin{tabular}{|C{2cm}||C{0.5cm}|C{0.5cm}|C{0.5cm}|C{0.5cm}|C{0.5cm}|C{0.5cm}|C{0.5cm}|C{2.5cm}|}
  \hline
  \textbf{Symmetry}&  $\mathbf{\overline{1}'}$  &  $\mathbf{m}$ & $\mathbf{2}$ & $\mathbf{m'}$ & $\mathbf{2'}$ & $\mathbf{3}$ & $\mathbf{\overline{3}'}$ & MPG \\
  \hline \hline
 $H=0$ & $+$ &  - & - & - & - &  + & + & $\overline{3}'$\\ \hline
  \multirow{2}{*}{$\mathbf{H}\| [001]$} & $\pm$ & +  & - & - & $\pm$ & - & - & $m\: (2'/m)$  \\ 
    & \cellcolor{Gray}$\pm$ & \cellcolor{Gray}+ & \cellcolor{Gray}$\pm$ & \cellcolor{Gray}+ & \cellcolor{Gray}+ & \cellcolor{Gray}-  & \cellcolor{Gray}- &  \cellcolor{Gray}$2'm'm \: (mm'm)$  \\ \hline
  \multirow{2}{*}{$\mathbf{H}\| [001]$} &  $\pm$ & + & -  & - & $\pm$ & - & - & $m\: (2'/m)$ \\
     &  \cellcolor{Gray}$\pm$ &  \cellcolor{Gray}+ & \cellcolor{Gray}{}$\pm$  & \cellcolor{Gray}+ &  \cellcolor{Gray}+ &  \cellcolor{Gray}- & \cellcolor{Gray}- &  \cellcolor{Gray}$m'm 2'\: (m'mm)$\\ \hline 
		\end{tabular}
  \end{center}
	\caption{The preserved symmetry elements and the corresponding MPGs of the calculated magnetic structure in the GS and in $H>0$ for $D>0$. $+$ and $-$ denote preserved and broken symmetries, respectively. $\pm$ denotes a symmetry element which is broken but might appear preserved in the experiment, with the corresponding MPGs given in parentheses \cite{suppl-mater}. Shaded cells represent results for $H \geq H_{c1}$.}
	\label{tab:symmetry}
\end{table}
%
%
%
\indent All MPGs in table \ref{tab:symmetry} allow the linear ME effect, while the polar MPGs in $H>0$ also allow field-induced ferroelectricity and bilinear ME effect \cite{Schmid-2008,Newnham,Bilbao-mtensor}. We focus here on the linear ME effect which seems to be the dominant contribution in Cu$_3$TeO$_6$. For the GS MPG $\overline{3}'$, tensor $\bm{\alpha}$ allows finite $P$ for any $H$ direction \cite{suppl-mater}, in disagreement with our results. For $\alpha$ in finite $H$ we refer to table \ref{tab:symmetry}. For $\mathbf{H}\|[001]$ and $D>0$ we have \cite{Schmid-2008,Newnham}
\begin{equation}\label{eq:alpha001}
	\bm{\alpha}_{m}=
	\begin{bmatrix}
0 & 0 & \alpha_{ac}\\
	0 & 0 & \alpha_{bc} \\
\alpha_{ca}& \alpha_{cb} & 0
	\end{bmatrix}, \;
 \bm{\alpha}_{2'm'm}=
	\begin{bmatrix}
0 & 0 & \alpha_{ac}\\
	0 & 0 & 0 \\
\alpha_{ca}& 0 & 0
	\end{bmatrix},
\end{equation}
which results in $P_{sat,a}=\alpha_{ac} H$ for all $H$. For $H||[010]$ and $D>0$ we have
\begin{equation}\label{eq:alpha010}
	\bm{\alpha}_{m}=
	\begin{bmatrix}
0 & \alpha_{ab} & 0\\
	\alpha_{ba} & 0 & \alpha_{bc} \\
0& \alpha_{cb} & 0
	\end{bmatrix}, \;
\bm{\alpha}_{m'm2'}=	
\begin{bmatrix}
0 & 0 & 0\\
	0 & 0 & \alpha_{bc} \\
0& \alpha_{cb} & 0
	\end{bmatrix},
\end{equation}
which gives $P_{sat,a}=\alpha_{ab} H$ for $H<H_{c1}$, and $P_{sat,a}=0$ for $H\geq H_{c1}$. The same conclusions apply to nonpolar MPGs listed in parentheses in table \ref{tab:symmetry}. Adding the nonlinear contributions allowed by symmetry results in equivalent conclusions regarding the polarization components (see SM for details). Therefore we conclude that the nonlinear behavior of $P_{sat}$ observed for $\mathbf{H}\|[010]$ (Fig.~\sref{fig2}{c}) is a consequence of the spin reorientation accompanied by the change of MPG. Our results are supported by a recent paper on the linear ME effect in Cu$_3$TeO$_6$ \cite{Tang-2023}.\\
%
%
\indent The static electric polarization and magnetic torque measurements combined with phenomenological simulations demonstrate that magnetic-field-induced spin reorientation accompanied by the change of magnetic point group, needs to be taken into account to understand the linear ME effect observed in Cu$_3$TeO$_6$. While the field-induced changes of MPG are reported in other systems, e.g. Cr$_2$O$_3$ \cite{Fiebig-1996}, the transition from the nonpolar AFM GS to polar field-induced state is not common and has been reported only in a few $4f-3d$ systems \cite{Shen-2019,Yanda-2021}. The mechanism of the ME effect in those systems relies on the interaction between the $4f$ and $3d$ magnetic ions. This cannot be applied to Cu$_3$TeO$_6$. Our symmetry analysis suggests that the calculated polar structures have weakly differing nonpolar counterparts in moderate magnetic fields, resulting in apparent linearity  of the ME effect in Cu$_3$TeO$_6$. The nonpolar to polar transition is supported by the strong spin-phonon coupling \cite{Caimi,Choi} and very slow AFM domain dynamics observed in Cu$_3$TeO$_6$ in weak magnetic field \cite{Herak-2005}, and the mechanism is probably rooted in the strong spin-lattice coupling which is not accounted for in our analysis. The change of magnetic symmetry in the applied magnetic field is critical to consider in further studies of the topological properties of Cu$_3$TeO$_6$ and similar topological antiferromagnets.
%
%
\newline
After the submission of this work, two papers reported linear ME effect in Cu$_3$TeO$_6$ \cite{Shahee-2023,Tang-2023}. The authors were not aware of the spin reorientation in nonzero magnetic field. Their results support the symmetry analysis presented in this work.
\begin{acknowledgments}
This work was supported by the Croatian Science Foundation (Grant IP-2018-01-2730) and projects Cryogenic Centre at the Institute of Physics - KaCIF (Grant KK.01.1.1.02.0012) and Centre for Advanced Research of Complex Systems CeNIKS (Grant KK.01.1.1.02.0013), co-financed by the Croatian Government and the European Union through the European Regional Development Fund - Competitiveness and Cohesion Operational Programme. V.~K. and D.~D. acknowledge funding of Croatian Science Foundation through grant HRZZ DOK-09-2018. X.~R. acknowledges funding from the French National Research Agency (ANR - Grant ANR-19-CE08-0013-02; HTHPCM Project) and GENCI for granting access to the HPC resources of [TGCC/CINES/IDRIS] under the allocation 2022-A0130907682. 
\end{acknowledgments}
%
%
%
%
\bibliographystyle{apsrev4-2}
\bibliography{main}
\end{document}